\pdfoutput=1

\documentclass[11pt]{article}

\usepackage[final]{acl}

\usepackage{times}
\usepackage{latexsym}
\usepackage[T1]{fontenc}
\usepackage[utf8]{inputenc}
\usepackage{microtype}
\usepackage{inconsolata}
\usepackage{multirow}
\usepackage{graphicx}
\usepackage{tikz}
\usepackage{makecell}
\usepackage{hyperref}
\usepackage{booktabs}
\usepackage{url}
\usepackage{arydshln}

\usepackage{xcolor}

\newcommand{\xmark}{%
\tikz[scale=0.23] {
    \draw[line width=0.7,line cap=round] (0,0) to [bend left=6] (1,1);
    \draw[line width=0.7,line cap=round] (0.2,0.95) to [bend right=3] (0.8,0.05);
}}
\newcommand{\cmark}{%
\tikz[scale=0.23] {
    \draw[line width=0.7,line cap=round] (0.25,0) to [bend left=10] (1,1);
    \draw[line width=0.8,line cap=round] (0,0.35) to [bend right=1] (0.23,0);
}}

\newcommand{\tmark}{%
\tikz[scale=0.18]{
  \draw[black, line width=0.8] (0,0) -- (0.5,1) -- (1,0) -- cycle;
}}

\title{WoW-Bench \raisebox{-0.3ex}{\includegraphics[height=1.6\ht\strutbox]{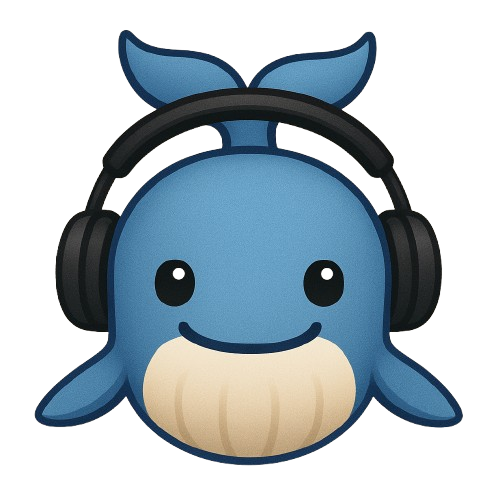}}: Evaluating Fine-Grained Acoustic Perception in Audio-Language Models via Marine Mammal Vocalizations}
\author{
  \quad\textbf{Jaeyeon Kim\textsuperscript{1}\thanks{This work was done at Seoul National University.}}
  \quad\textbf{Heeseung Yun\textsuperscript{2}}
  \quad\textbf{Sang Hoon Woo\textsuperscript{2}} \\
  \quad\textbf{Chao-Han Huck Yang\textsuperscript{3}} 
  \quad\textbf{Gunhee Kim\textsuperscript{2}} \\
  \textsuperscript{1} Carnegie Mellon University
  \quad \textsuperscript{2} Seoul National University \quad \textsuperscript{3}NVIDIA \\
  {\small \texttt{jaeyeon2@andrew.cmu.edu, heeseung.yun@vision.snu.ac.kr, tonyswoo@gmail.com}} \\
  {\small \texttt{hucky@nvidia.com, gunhee@snu.ac.kr}}
}

\begin{document}
\maketitle
\begin{abstract}
Large audio-language models (LALMs) extend language understanding into the auditory domain, yet their ability to perform low-level listening, such as pitch and duration detection, remains underexplored. 
However, low-level listening is critical for real-world, out-of-distribution tasks where models must reason about unfamiliar sounds based on fine-grained acoustic cues.
To address this gap, we introduce the World-of-Whale benchmark (WoW-Bench) to evaluate low-level auditory perception and cognition using marine mammal vocalizations. We use marine mammal vocalizations as out-of-distribution sound events to better assess models' low-level listening and so that the models do not rely on prior knowledge of the sound events. WoW-bench is composed of a Perception benchmark for categorizing novel sounds and a Cognition benchmark, inspired by Bloom’s taxonomy, to assess the abilities to remember, understand, apply, and analyze sound events. For the Cognition benchmark, we additionally introduce distractor questions to evaluate whether models are truly solving problems through listening rather than relying on other heuristics. Experiments with state-of-the-art LALMs show performance far below human levels, indicating a need for stronger auditory grounding in LALMs.\footnote{Project page: \url{https://jaeyeonkim99.github.io/wow_bench/}}

\end{abstract}

\section{Introduction}
Humans naturally perceive, interpret, and reason about sound events in their environment. 
Motivated by the success of large language models (LLMs) in natural language understanding and reasoning, recent works have developed large audio-language models (LALMs) that expand these capabilities to the auditory domain \cite{ltu, salmonn, gama, qwen_audio_1, qwen_llm}. These models integrate an audio encoder with an LLM to support general audio understanding and instruction following across diverse sound-related tasks.
A number of benchmarks have been introduced to evaluate and advance LALMs' auditory understanding and reasoning, including compositional~\cite{compa}, deductive~\cite{audio_entail}, and comparative~\cite{adiff} reasoning.
More recently, broader benchmarks have been proposed to evaluate a wider spectrum of auditory comprehension and reasoning skills \cite{mmau}.

\begin{figure*}[t]
\centering
\includegraphics[width=\textwidth]{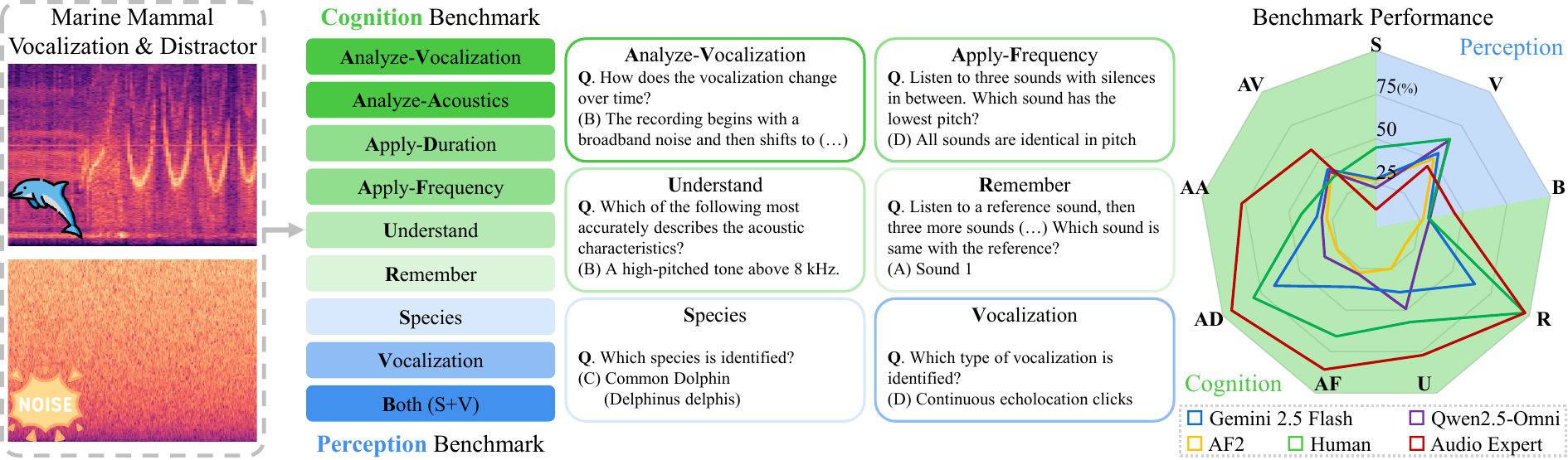}
\vspace{-20pt}
\caption{\textbf{World-of-Whale} benchmark aims to evaluate low-level listening capabilities of LALMs using marine mammal vocalizations, which are rarely represented in conventional datasets and span a broad acoustic range. LALMs struggle with Cognition questions that humans can reliably solve by using low-level auditory perception.}
\label{fig:dataset_overview}
\vspace{-10pt}
\end{figure*}

However, the perceptual capability that \textit{precedes} the reasoning of LALMs' auditory input remains relatively underexplored, despite its foundational importance for auditory understanding and reasoning. This gap is critical, as perceptual errors often account for a substantial portion of failures in reasoning tasks \cite{mmau}.
Current evaluations of LALMs’ perceptual capabilities predominantly rely on audio classification tasks \cite{audioset, vggsound, esc50, airbench}, which map acoustic signals to predefined semantic labels such as “dog barking” or “siren.” While these benchmarks assess a model’s ability to perform semantic categorization, they offer limited insight into whether models are attending to fine-grained acoustic features, such as pitch and duration. Humans naturally process these acoustic cues prior to deriving semantic meaning \cite{auditory_scene}.

These low-level acoustic features are essential for forming auditory events and objects \cite{auditory_scene, psychology_hearing}. Focusing on this perspective, we define \textit{low-level listening} as the perceptual ability to detect and differentiate elementary acoustic attributes prior to semantic interpretation or categorization \cite{auditory_scene, psychology_hearing, low-high}.
Low-level listening also plays a critical role in understanding and reasoning in unfamiliar scenarios, particularly when combined with knowledge about the acoustic characteristics of novel events. Real-world applications such as industrial anomaly detection \cite{anomaly_2020, anomaly_2022}, auditory surveillance systems \cite{crocco2016audio}, and bioacoustic monitoring \cite{bardeli2010detecting} require strong out-of-distribution (OOD) generalization, as these tasks involve rare and anomalous sounds often absent from training data~\cite{anomaly_2020}.
Effective models should detect such unencountered anomalies by combining knowledge about patterns and low-level acoustic cues such as high-frequency whines.

Recognizing the importance and the relatively overlooked nature of low-level listening, we propose the \textbf{World-of-Whale Benchmark (WoW-Bench)} to assess the fine-grained perceptual capabilities of LALMs. Our benchmark focuses on marine mammal vocalizations, which span a broad acoustic range from below 20 Hz to over 20 kHz, and exhibit diverse vocalization patterns such as click, whistle, and calls. Moreover, these vocalizations are rarely covered in conventional large-scale audio corpora, \textit{e.g.}, only one out of 527 AudioSet labels.
By evaluating models in this underrepresented and acoustically rich out-of-domain setting, we aim to provide a more rigorous and fine-grained test of perceptual ability in LALMs.

Our benchmark is structured into two components. First, we assess the perceptual generalization of LALMs by evaluating their ability to categorize sounds into less familiar classes based on low-level listening. Second, we examine their cognitive processing ability, focusing on how well models can interpret and decompose information gained through low-level auditory perception.
Inspired by the taxonomy of cognitive hierarchy for learning new concepts~\cite{bloom_original, bloom_revised}, we systematically evaluate models’ abilities to \textit{remember}, \textit{understand}, \textit{apply}, and \textit{analyze} auditory information, as illustrated in Figure~\ref{fig:dataset_overview}.
Additionally, we introduce distractor questions to evaluate whether models are truly solving tasks through listening rather than relying on shallow heuristics or linguistic priors.

Extensive experiments with existing LALMs reveal that both their perceptual generalization and cognitive processing based on low-level listening are limited. As shown in Figure~\ref{fig:dataset_overview}, even the best model achieves significantly lower results than humans on the Cognition tasks, highlighting substantial room for improvement in the low-level listening capabilities of LALMs. Moreover, qualitative analysis shows that models tend to adopt a classify-first strategy and infer acoustic properties based on presumed categories rather than listening to them, which can lead to incorrect decisions.

\section{Related Work}
\noindent\textbf{Large Audio-Language Models.}
Recent advances in LALMs have substantially improved performance on tasks requiring understanding and reasoning over general sound events. Pengi \cite{pengi} is among the first to unify diverse audio tasks under a single text generation framework by connecting an audio encoder to a decoder-only language model, achieving strong results across a wide range of downstream tasks. Subsequently, a number of LALMs have been introduced \cite{ltu, ltu_as, salmonn, af1, qwen_audio_1, qwen2_audio, gama}, aligning pretrained audio encoders with large language models and training on large-scale audio-text datasets. These models can follow language instructions and perform a wide range of audio tasks, demonstrating strong performance on both closed-ended tasks (\textit{e.g.}, audio classification) and open-ended tasks (\textit{e.g.}, audio captioning and QA).

More recent efforts focus on enhancing the reasoning capabilities of LALMs by introducing carefully designed training corpora \cite{gama, af2, af3, mellow, audio_reasoner} or incorporating chain-of-thought prompting \cite{cot, audio_reasoner, audio_cot}.
While these approaches advance high-level reasoning, comparatively little attention has been paid to low-level auditory perception of the models which serves as a foundation for robust reasoning and understanding \cite{mmau}. In this work, we address this gap by proposing a new benchmark specifically designed to assess the low-level listening abilities of state-of-the-art LALMs.

\noindent\textbf{Benchmarking Large Audio-Language Models. }
Following the rapid development of LALMs, several benchmarks have been proposed to assess their capabilities across diverse understanding and reasoning abilities. These benchmarks evaluate compositional reasoning over complex sound events~\cite{compa, gama}, temporal reasoning over multiple sound events~\cite{trea}, deductive reasoning via textual entailment conditioned on audio inputs~\cite{audio_entail}, and long-context understanding using long audio segments \cite{af2}. Some benchmarks evaluate LALMs across a broad range of domains, including speech, vocal sounds, general audio, and music \cite{audiobench, airbench, mmau, mmar}. Among these, AIRBench~\cite{airbench} and AudioBench~\cite{audiobench} primarily focus on audio understanding and instruction-following tasks, while MMAR~\cite{mmar} focuses on multi-step reasoning within the general sound event domain. 

However, these benchmarks primarily target reasoning over the perceived sound events, while low-level listening abilities, which are how the model actually listens to and interprets acoustic input, remain largely unexplored. Among existing works, ADIFF~\cite{adiff} and MMAU~\cite{mmau} place relatively greater emphasis on perception. ADIFF addresses comparative reasoning by prompting models to describe the differences between audio clips, while MMAU focuses on information extraction and reasoning across diverse audio types. In contrast, our benchmark is specifically designed to evaluate how LALMs perceive and process novel acoustic events and fine-grained auditory details, thereby assessing their low-level listening capabilities.

\section{World-of-Whale Benchmark}
\begin{table*}
\centering
\resizebox{\textwidth}{!}{
\begin{tabular}{llccl}
\hline
\multirow{2}{*}{Dataset} & \multirow{2}{*}{Target Ability} & \multirow{2}{*}{\makecell[l]{Low-level\\Listening}} & \multirow{2}{*}{Semantic Source} & \multirow{2}{*}{\#Test} \\
                         &                           &                                     &                                 &                             \\
\hline
CompA \citeyearpar{compa}         & Compositional reasoning & \xmark & Casual, Synthetic & 0.6k \\
CompA-R-\textit{test} \citeyearpar{gama}         & Complex reasoning & \xmark & Casual  & 1.6k \\
MAE \citeyearpar{mae} & Multi-audio processing & \xmark & Casual & 9k \\ 
AIRBench \citeyearpar{airbench}    & Audio understanding \& Instruction following & \xmark & Casual & 4.5k$^*$ \\
AudioBench \citeyearpar{audiobench}& Audio understanding \& Instruction following & \xmark & Casual & 8.9k$^*$ \\
Audio Entailment \citeyearpar{audio_entail} & Deductive reasoning & \xmark & Casual & 5.8k \\
ADIFF \citeyearpar{adiff}          & Comparative reasoning & \tmark & Casual & 10k \\
MMAU \citeyearpar{mmau}  & Information extraction \& Reasoning & \xmark & Casual, Synthetic & 3.3k$^*$ \\
LongAudioBench \citeyearpar{af2}   & Long context understanding & \xmark & Casual, Egocentric & 2.4k \\
TREA \citeyearpar{trea} & Temporal reasoning & \xmark & Casual & 0.6k \\
MMAR \citeyearpar{mmar} & Multi-step reasoning & \xmark & Casual & 0.4k$^*$ \\

\cline{1-5}
WoW-Bench &
Perceptual generalization \& Cognition based on low-level listening &
\cmark &
Marine Mammal \citeyearpar{watkins} &
1.7k
\\
\hline
\end{tabular}
}
\vspace{-7pt}
\caption{
Comparison to existing LALM benchmarks. \textit{Low-level Listening} indicates that the dataset evaluates a model’s ability to process fine-grained acoustic attributes prior to semantic categorization. \protect\tmark\ indicates partial assessment of the perceptual capabilities like low/high pitch.
\textit{Casual} indicates the dataset corpora covers general sound events like human and music, \textit{e.g.}, \citet{audioset,audiocaps,clotho,mira}.
$^*$ denotes the number of general audio-related questions in each test set.
}
\label{tab:comparison}
\vspace{-10pt}
\end{table*}

\subsection{Overview}
We introduce the \textbf{World-of-Whale Benchmark (WoW-Bench)}, designed to evaluate LALMs on their ability to perceive and cognitively process low-level acoustic details in unfamiliar and acoustically diverse scenarios. The benchmark has two components:
(1) The \textbf{Perception} benchmark that tests the model’s ability to categorize unfamiliar sound events based on low-level listening and their internal knowledge (\S\ref{sec:perception}), and
(2) The \textbf{Cognition} benchmark that assesses whether models can cognitively process fine-grained acoustic characteristics and perceived events through low-level listening, as illustrated in Figure~\ref{fig:dataset_overview}~(\S\ref{sec:cognition}). We further describe the design of adversarial distractors used in the Cognition benchmark in \S\ref{sec:distractor}, which test whether models are truly listening to solve the questions.

WoW-Bench is the first benchmark to explicitly focus on low-level listening and perceptual ability in a controlled OOD setup. 
As shown in Table~\ref{tab:comparison}, prior benchmarks often rely on widely used datasets \cite{audioset, audioset_strong, audiocaps, clotho} or content collected from similar in-the-wild videos \cite{mira, video_recap}, i.e., featuring similar distribution of sound events.
However, we observe that LALMs often bypass low-level acoustic details and prioritize semantic interpretation of sound based on in-domain knowledge to infer acoustic characteristics (e.g. "pig oinking" in Figure~\ref{fig:qual1}-(b)).
As illustrated in Figure~\ref{fig:range}-(a), when we construct similar questions with in-domain ESC-50 audio \cite{esc50}, model performance is consistently higher compared to OOD, but the gap largely varies across tasks.
Such variance demonstrates the difficulty of disentangling whether models \textit{know} the sound from familiar domains a priori or actually \textit{listen} to it.
To provide a more controlled perspective, we believe OOD inputs are crucial in evaluating low-level listening, by reducing the influence of acoustic knowledge about the classified sound (See \S\ref{appen:esc50}).

To this end, WoW-Bench utilizes marine mammal vocalizations, which are underrepresented in common corpora.
For example, AudioSet contains only one relevant label, VGGSound \cite{vggsound} contains two, and both ESC-50 \cite{esc50} and FSD50K \cite{fsd50k} contain no relevant labels.
This approach allows for the evaluation of whether models can generalize beyond familiar data and attend to acoustic content, rather than relying solely on training priors.
Identifying suitable OOD audio content is itself a non-trivial challenge, as the pretraining data of many LALMs already encompass a wide range of everyday sounds.
We observed that even modest extensions, such as adding other animal vocalizations from the VGGSound test set, can reduce the OOD nature of the task.
For instance, when asked to classify animal species, Qwen2-Audio-Instruct's accuracy increased markedly from \textbf{28.3\%} on WoW-Bench to \textbf{76\%} on the VGGSound-based variant (e.g., distinguishing Baltimore oriole from pigeon).

WoW-Bench consists entirely of multiple-choice questions (MCQs) in light of two observations. While humans may struggle to generate accurate descriptions of unfamiliar sounds, they can often identify the correct option by comparing acoustic details across choices using low-level listening. Moreover, MCQs enable standardized evaluation and are widely adopted for both LLM~\cite{mmlu, mmmu} and LALM benchmarking~\cite{airbench, mmau}.

\subsection{Perception Benchmark}\label{sec:perception}
The Perception benchmark is designed to judge whether models can classify audio events with unfamiliar labels and acoustic characteristics, based on low-level listening abilities and prior knowledge regarding the problem domain. It measures the perceptual generalization of LALMs to novel domains, where successful classification depends not on memorized patterns but on the model’s ability to listen and interpret subtle acoustic features.

\noindent
(1) The \textbf{Species} task requires models to classify each audio clip of a vocalization into the correct species category, such as humpback whale, killer whale, or melon-headed whale. This task is analogous to typical audio event classification and has been a central focus in bioacoustics research. It demands fine-grained auditory discrimination, as vocalization differences between species are often subtle. 
In our benchmark, we adopt a zero-shot setting, requiring models to map fine-grained perceptual cues to species-level knowledge, if such knowledge exists within the model.

\noindent
(2) The \textbf{Vocalization} task involves models selecting the most appropriate natural language description for a given vocalization. These descriptions refer to vocalization types such as clicks, whistles, or calls, and may also include acoustic characteristics like “high-pitched” or “contains background ship noise.” Similarly to the Species task, the Vocalization task requires models to capture low-level acoustic details and map them to unfamiliar labels. However, the vocalization labels are often more intuitive, semantically interpretable, and directly related to the acoustic signal than the name of species. As a result, this task relies more on perceptual matching than on prior knowledge.

\noindent
(3) The \textbf{Both} task requires models to select the option that correctly describes both the species and the vocalization type for a given audio clip, for example,``Leopard Seal – Long Call.'' This task presents a more challenging scenario by combining two subtasks and assessing compositional auditory perception. In this setup, models must capture different dimensions of acoustic information simultaneously to perform successfully.

\subsection{Cognition Benchmark}\label{sec:cognition}
Motivated by Bloom’s taxonomy~\cite{bloom_original, bloom_revised}, which defines a cognitive hierarchy of learning objectives, we design the Cognition benchmark comprising four subtask types to evaluate different aspects of how LALMs process low-level acoustic details and perceived audio events.

\noindent
(1) The \textbf{Remember} level in the hierarchy involves recognizing and recalling previously encountered information. In this task, the model is given a reference sound, followed by three additional sounds separated by silence. The model must identify which of the subsequent sounds is identical to the reference. This tests the model's ability to recall the reference and recognize similarity based on acoustic characteristics across the segments, without relying on the understanding of sound events.

\noindent
(2) In Bloom's taxonomy, the \textbf{Understand} level refers to the ability to interpret observed information. In our benchmark, we evaluate whether a model comprehends the underlying acoustic properties of a sound by selecting the description that most accurately represents the low-level acoustic features. To reduce ambiguity, each choice includes both a perceptual pitch label and the corresponding frequency range, \textit{e.g.}, "upsweeping high-frequency tonal sound above 8 kHz."

\noindent
(3) The \textbf{Apply} level in the taxonomy denotes the capacity to utilize acquired knowledge in new situations. Accordingly, our Apply task evaluates whether the model can transfer low-level listening abilities to novel contexts. Specifically, the evaluation consists of two parts; in each part, the model compares audio clips based on a core acoustic property, either pitch or duration. In the pitch-based subtask (\textit{Apply–Frequency}), the model receives three clips and identifies the one with either the highest or lowest pitch. In the duration-based subtask (\textit{Apply–Duration}), the model selects the sound with the longest or shortest duration. This task requires the model to utilize low-level perceptual information in a comparative setting, without engaging semantic understanding of the individual sounds.

\noindent
(4) Finally, the \textbf{Analyze} level in Bloom's taxonomy involves breaking information into parts and examining their relationships.
To assess this cognitive skill, we investigate whether the model can analyze transitions within complex acoustic sequences. In particular, we present the model with an audio clip composed of two sounds without any intervening silence and the model must identify and interpret the transition between the sounds. We further divide the task into two subtasks: (i) \textit{Analyze–Acoustics}, which focuses on identifying transitions based on low-level acoustic cues, such as a shift from a low-frequency tone to a broadband pulse; and (ii) \textit{Analyze–Vocalization}, which involves detecting changes in higher-level auditory characteristics like species and vocalization types, such as a transition from a Beluga’s squeal to a Common Dolphin’s whistle. We carefully construct the answer choices so that attending to only one part of the sequence is insufficient, requiring the model to analyze the shift between segments.

\subsection{Distractors and QA Robustness}\label{sec:distractor}
\newcite{mmau} replace audio inputs with random noise and compare the performance with that on real audio to assess whether models genuinely attend to audio or rely primarily on language priors. 
However, we believe it is more informative to observe how the model’s prediction changes in response to altered audio input, rather than simply verifying the correctness of responses.

Building on this perspective, we develop a more targeted approach by introducing adversarial distractors tailored to each Cognition question type. Each distractor incorporates audio that is qualitatively distinct from those of regular questions for the same type, such that the less expected choice within the context becomes the correct answer, aiming to directly test whether the model is truly solving the task through listening.
For example, in the \textit{Apply–Frequency} task, where the model is asked to select the sound with the highest pitch (A. Sound 1, B. Sound 2, C. Sound 3, D. All indistinguishable), we present three identical sounds as input (\textit{i.e}., the distractor audio). In such case, the model may still choose one of the labeled sounds, as option D can be less expected given the question format. This approach reveals whether the model relies on shallow heuristics or demonstrates genuine listening abilities, thereby providing a more rigorous test of the model's perceptual abilities.

We generate distractor variants by selecting a subset of non-distractor questions and replacing their audio components with distractor audio. We systematically design the distractors by inverting the expected pattern for each question type, thereby encouraging the model to rely on listening to solve the task. For instance, for question types that require comparison across multiple audio samples, we reverse the common expectation that one of the labeled candidate sounds matches the correct answer. Therefore, in the Remember question type, all candidates differ from the reference, making “None of the above” the correct choice. Further details are provided in the Appendix~\ref{appen:distractor_design}.

\subsection{Dataset Curation}
All audio samples are sourced from the Watkins Marine Mammal Sound Database~\cite{watkins}. Except for manually constructed Remember questions, all other questions are generated using a large language model and subsequently undergo rigorous human verification and reannotation. For each Perception task, we generate 100 questions. For each Cognition task, including subtasks within Apply and Analyze, we generate 200 questions, along with 50 adversarial distractor questions per task before filtering. As a result, the final benchmark comprises \textbf{1,777} validated questions, forming a test-only dataset designed for zero-shot evaluation. A detailed description of the curation process and dataset statistics is provided in the Appendix~\ref{appen:data_curation}.
\begin{table*}[t]
\centering
\resizebox{\textwidth}{!}{
\begin{tabular}{l|c|cccc|ccccccccc|c|c}
\hline
\multirow{2}{*}{\textbf{Model}} 
  & \multirow{2}{*}{\textbf{Params}} 
  & \multicolumn{4}{c|}{\textbf{Perception}} 
  & \multicolumn{9}{c|}{\textbf{Cognition}} 
  & \multirow{2}{*}{\textbf{Total}} 
  & \multirow{2}{*}{\textbf{MMAU}} \\
  & 
  & \textbf{S} & \textbf{V} & \textbf{B} & \textbf{All} 
  & \textbf{R} & \textbf{U} & \textbf{AF} & \textbf{AD} & \textbf{AA} & \textbf{AV} & \textbf{ND} & \textbf{D} & \textbf{All} 
  &  &  \\
\hline
Random & -  & 27.3  & 22.7 & 22.0 & 24.0 & 24.8 & 25.8 & 22.0 & 23.6 & 28.0 & 25.3 & 26.6  & 23.3 & 24.9 & 24.8 & -- \\
SALMONN \citeyearpar{salmonn} & 13B & 24.2 & 14.4 & 8.0 & 15.5 & 26.8 & 18.8 & 32.0 & 26.4 & 0.0 & 3.3 & 21.9 & 2.7 & 18.0 & 17.6 & 42.1 \\
LTU \citeyearpar{ltu} & 7B & \underline{28.3} & 29.9 & 24.0 & 27.4 & 22.8 & 30.8 & 8.4 & 20.4 & 22.0 & 32.0 & 19.2 & 36.0 & 22.6 & 23.4 & 20.7 \\
LTU-AS \citeyearpar{ltu_as} & 7B & 24.2 & 20.6 & 22.0 & 22.3 & 20.0 & 23.3 & 18.0 & 22.0 & 21.6 & 26.6 & 24.3 & 12.3 & 21.9 & 22.0 & -- \\
GAMA \citeyearpar{gama} & 7B & 21.2 & 29.9 & 24.0 & 25.0 & 23.6 & 22.5 & 21.6 & 22.0 & 20.4 & 23.2 & 14.8 & \underline{51.3} & 22.2 & 22.7 & 30.7 \\
GAMA-IT \citeyearpar{gama} & 7B & 19.2 & 24.7 & 22.0 & 22.0 & 19.6 & 25.0 & 18.0 & 18.8 & 26.8 & 24.5 & 12.3 & \textbf{60.7} & 22.1 & 22.1 & 32.7 \\
Qwen-Audio-Chat \citeyearpar{qwen_audio_1} & 8.4B & 24.2 & 36.1 & 19.0 & 26.4 & 30.0 & 30.4 & 40.0 & 24.0 & \underline{33.2} & 34.0 & 38.9 & 4.67 & 31.9 & 31.0 & -- \\
Qwen2-Audio-Instruct \citeyearpar{qwen2_audio} & 8.4B & \underline{28.3} & 41.2 & 19.0 & 29.4 & 19.2 & 30.0 & 33.2 & 25.2 & 27.6 & 44.8 & 31.0 & 25.7 & 29.9 & 29.8 & 61.2 \\
Qwen2.5-Omni \citeyearpar{xu2025qwenomni} & 10.7B & 22.2 & \underline{63.9} & 31.0 & \underline{38.9} & 28.8 & \textbf{49.2} & 28.4 & 33.6 & 31.2 & 41.1 & 37.4 & 26.7 & 35.3 & 35.9 & \textbf{76.8} \\
Audio Flamingo 2 \citeyearpar{af2} & 3.3B & 26.3 & 50.5 & 27.0 & 34.5 & 19.2 & 25.0 & 27.6 & 25.6 & 28.4 & 39.4 & 29.2 & 20.7 & 27.5 & 28.6 & 68.1 \\
Audio Flamingo 3 \citeyearpar{af3} & 8.2B & \textbf{54.6} & \textbf{75.3} & \textbf{58.0} & \textbf{62.5} & 23.6 & 33.8 & 38.4 & 24.0 & 31.6 & \underline{50.6} & 38.4 & 14.7 & 33.6 & 38.4 & \underline{75.8} \\
Gemini 1.5 Pro & -- & 23.2 & 35.1 & 18.0 & 25.3 & 28.4 & 26.7 & 26.0 & 30.0 & 30.0 & 34.0 & 35.1 & 5.7 & 29.2 & 28.5 & -- \\
Gemini 2.0 Flash & -- & 18.2 & 46.4 & 18.0 & 27.4 & 43.6 & 44.6 & \underline{42.4} & 26.8 & \underline{33.2} & 36.5 & 38.5 & 35.0 & 37.8 & 36.1 & 68.9 \\
Gemini 2.5 Flash & -- & 27.3 & 54.6 & 30.0 & 37.2 & \underline{64.4} & 39.2 & 36.0 & \underline{66.4} & \textbf{34.0} & 42.7 & \underline{54.5} & 18.3 & \underline{47.2} & \underline{45.5} & 69.5 \\
Gemini 2.5 Pro & -- & 16.2 & 59.8 & \underline{36.0} & 37.2 & \textbf{86.4} & \underline{48.3} & \textbf{59.6} & \textbf{86.8} & \textbf{34.0} & \textbf{52.3} & \textbf{65.1} & 46.7 & \textbf{61.4} & \textbf{57.3} & 70.6 \\
\hline
\end{tabular}}
\vspace{-7pt}
\caption{
Evaluation results on the WoW-Bench.
Each acronym stands for \textbf{S}pecies, \textbf{V}ocalization, \textbf{B}oth (Perception), \textbf{R}emember, \textbf{U}nderstand, \textbf{A}pply-\textbf{F}requency, \textbf{A}pply-\textbf{D}uration, \textbf{A}nalyze-\textbf{A}coustics, \textbf{A}nalyze-\textbf{V}ocalization, \textbf{N}on-\textbf{D}istractor, \textbf{D}istractor (Cognition), respectively.
Last two columns report the overall score and their performance on the Sound Test subset from MMAU-v05.15.25~\citeyearpar{mmau} as reference.
For each column, the highest score is indicated in \textbf{bold}, while the second-highest score is \underline{underlined}.
}
\vspace{-10pt}
\label{tab:main_results}
\end{table*}

\section{Experiment}

\subsection{Setup}
\noindent\textbf{Models.}
We evaluate a range of LALMs capable of processing non-speech sound events, including LTU~\cite{ltu}, LTU-AS~\cite{ltu_as}, SALMONN~\cite{salmonn}, GAMA, and GAMA-IT~\cite{gama}, as well as two instruction-following models from the Qwen-Audio series, Qwen-Audio-Chat~\cite{qwen_audio_1} and Qwen2-Audio-Instruct, on our proposed benchmark. We also consider Audio Flamingo 2~\cite{af2} and Audio Flamingo 3~\cite{af3}, recognized for their strong reasoning and long-context comprehension, and Qwen2.5-Omni~\cite{xu2025qwenomni}, a multimodal model which exhibits promising performance on audio tasks.
For commercial LALMs, we evaluate four models from the Gemini series, specifically Gemini 1.5 Pro~\cite{gemini1.5}, Gemini 2.0 Flash~\cite{gemini}, Gemini 2.5 Flash, and Gemini 2.5 Pro~\cite{gemini2.5}, multimodal models known for their strong general audio processing capabilities.

\noindent\textbf{Evaluation Strategy.} 
We report the micro-averaged accuracy across all questions in the benchmark, as well as each task and subtask. To account for varying instruction-following capabilities of different LALMs, we experiment with multiple prompting strategies, \textit{e.g.}, “choose the correct option,” “return the answer letter”, and report the score of the best-performing prompt for each model. Since LALMs vary in their response formats and do not consistently adhere to instructions regarding formatting, we use GPT-4.1-mini to extract the final answer and determine its correctness. More details of the evaluation prompts and answer extraction are provided in Appendix~\ref{appen:eval_detail}.

\subsection{Results}\label{sec:results}

The main evaluation results are presented in Table~\ref{tab:main_results}. Overall, all models perform poorly on the WoW-Bench. Even the best performing commercial model, Gemini 2.5 Pro, remains below 60\% accuracy. Most open-source models, except Qwen2.5-Omni and Audio Flamingo 3, perform at levels similar to random chance. These findings highlight significant limitations in both low-level listening and the associated cognitive processing capabilities of LALMs.
\begin{figure}[t]
\centering
\includegraphics[width=\linewidth]{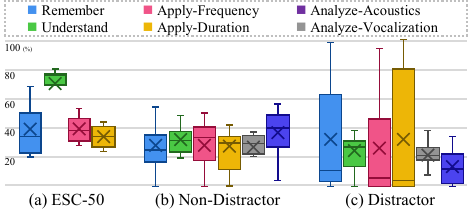}
\caption{
Performance distribution across cognition question types, grouped by the presence of dataset and distractor condition.
Each bar represents the interquartile range (Q1 to Q3) of model performance. The three horizontal lines of each bar correspond to the first quartile (Q1), median (Q2), and third quartile (Q3).
The × marker indicates the mean.
}
\label{fig:range}
\vspace{-10pt}
\end{figure}
\begin{figure}[t]
\centering
\includegraphics[width=\linewidth]{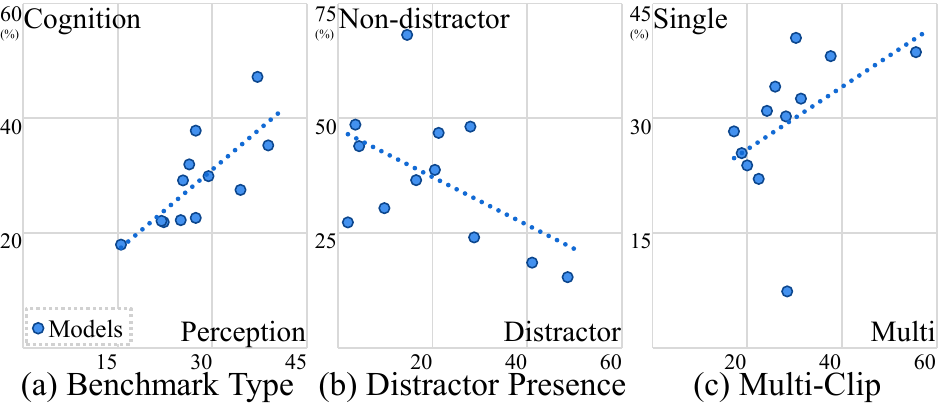}
\caption{
Distribution of models' performance regarding (a) benchmark type, (b) distractor presence, and (c) multi-clip, where we observe the Pearson correlation of 0.75, -0.51, and 0.47, respectively.
}
\label{fig:scatter}
\vspace{-10pt}
\end{figure}

\noindent\textbf{Q1. How do models perform across different tasks?}
One might intuitively assume that the Cognition tasks would be easier than the Perception tasks, as they do not require prior knowledge of marine mammals and can be addressed by simply analyzing the provided audio clip. However, this assumption does not consistently hold true. Aside from the Gemini series models, many models perform worse on the seemingly simpler Cognition tasks. 
For example, Audio Flamingo 2, one of the strong models on MMAU, achieves only 19.2\% accuracy on the Remember task, lower than its performance on the Species task (26.3\%) and even below the random baseline (24.8\%). This suggests that audio-grounded cognitive processing remains severely underdeveloped in current LALMs.

In the Perception benchmark, models typically exhibit higher performance on the Vocalization task, where labels such as “whistle” or “clicks” are more intuitively associated with the acoustic signal. The performance on the Species and Both tasks is close to random chance, indicating that models either lack relevant species-level knowledge or fail to capture the nuanced acoustic details necessary for accurate differentiation.

The results on the Cognition benchmark show a more diverse pattern, as illustrated in Figure~\ref{fig:range}. The performance varies dramatically across both tasks and models; while some models score near zero on certain tasks, others exceed 50\% accuracy. This variability highlights the strengths and weaknesses of each model’s low-level listening ability. For instance, Gemini 2.5 Flash achieves over 60\% accuracy on the Remember and Apply–Duration tasks, both of which involve comparing multiple audio segments. However, its performance drops to 36\% on Apply–Frequency, which involves similar comparative analysis, revealing a relative weakness in processing pitch-based information.

It is worth noting that some LALMs struggle with multi-audio comparison \cite{mae}. Their poor performance on the Cognition tasks that require multi-clip processing can therefore be attributed not only to limitations in low-level listening but also to architectural and comparative reasoning constraints. Nevertheless, performance on single-clip and multi-clip tasks shows a positive correlation as shown in Figure~\ref{fig:scatter}-(c), suggesting that the performance is still largely driven by shared low-level listening capabilities. Also, models do not consistently perform worse on multi-clip tasks, rather outcomes are more strongly influenced by the properties of each task and the intrinsic capabilities of the model. Note that our design is in line with the existing practice of using multiple concatenated events as a single audio recording~\cite{airbench, trea}.

Lastly, as shown in Figure~\ref{fig:scatter}-(a), models' performance on the Perception and Cognition tasks in WoW-Bench exhibits positive correlation, suggesting that both tasks rely on common low-level listening capabilities. For comparison, we also include their performance on MMAU Sound Test set in Table~\ref{tab:main_results}. Notably, a high score on MMAU does not necessarily translate to strong performance on WoW-Bench. For example, Audio Flamingo 2 and SALMONN perform worse than some models on WoW-Bench, despite outperforming them on MMAU. This suggests that WoW-Bench introduces new challenges and evaluation criteria not captured by existing benchmarks.

\begin{figure}[t]
\centering
\includegraphics[width=0.95\linewidth]{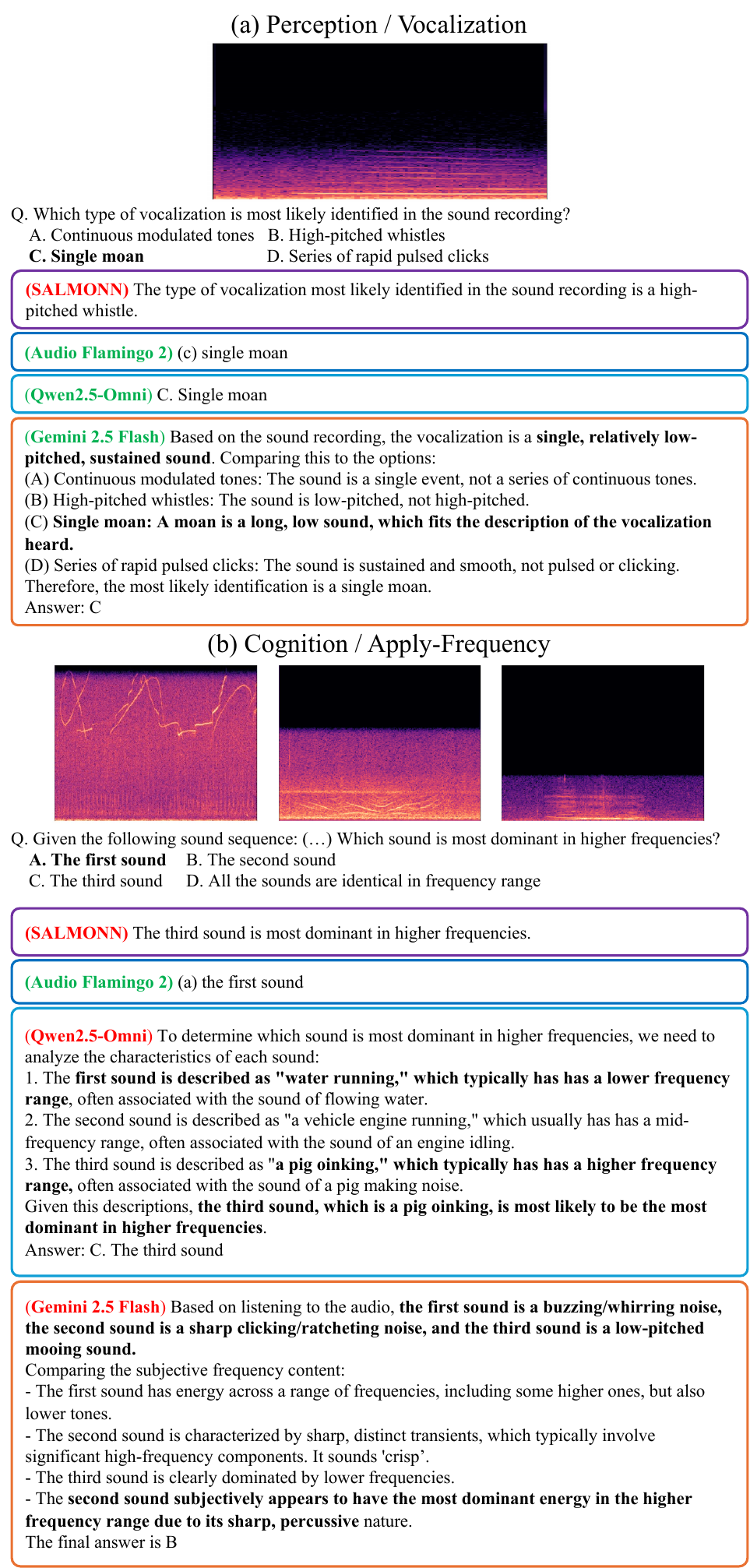}
\vspace{-5pt}
\caption{
Qualitative examples of model responses to (a) a Perception/Vocalization question and (b) a Cognition/Apply-Frequency question.
}
\label{fig:qual1}
\vspace{-15pt}
\end{figure}

\begin{figure}[t]
\centering
\includegraphics[width=\linewidth]{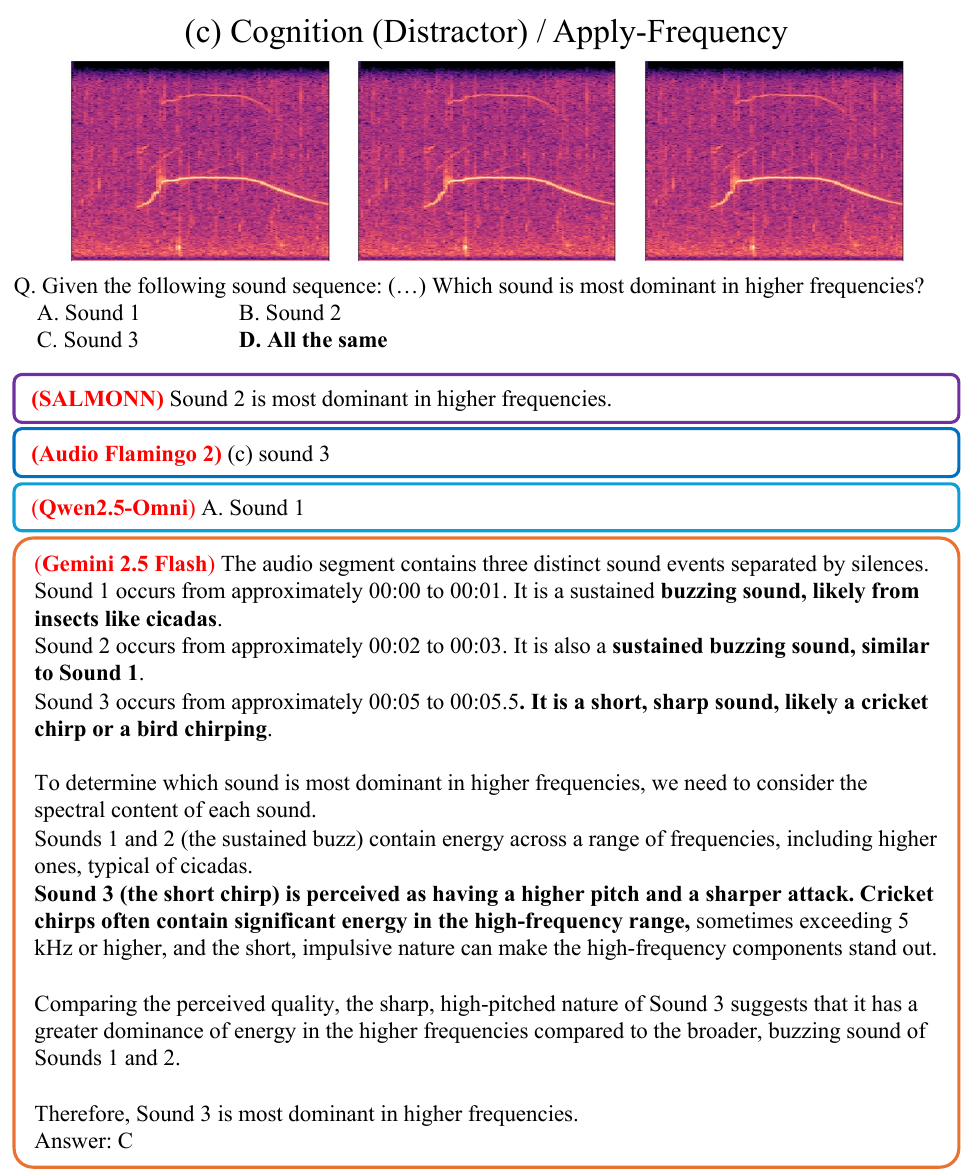}
\caption{
Qualitative example of a model response to (c) a distractor question for Cognition/Apply-Frequency task.
}
\label{fig:qual2}
\vspace{-10pt}
\end{figure}

\noindent\textbf{Q2. How do humans perform?}
As shown in Figure~\ref{fig:dataset_overview}, human and model performance are comparable in the Perception benchmark, as humans often struggle due to limited prior knowledge about marine mammals. However, in the Cognition tasks, humans significantly outperform the models; for example, they achieve 97.1\% on the Remember task, while the Gemini 2.5 Flash model achieves only 57.1\%. These results demonstrate that the questions are well-constructed and reliable; they assess features that humans can readily identify through their low-level listening ability even in an unfamiliar domain, yet remain challenging for current LALMs. 

To further evaluate the benchmark’s diagnostic value, we include listeners with strong backgrounds in audio signal processing. Expert listeners consistently outperform inexperienced participants in Cognition tasks. This indicates that our benchmark tasks effectively capture cognitive processing of fine-grained acoustic properties. Details of the human evaluation are provided in Appendix~\ref{appen:human_eval}.

\noindent\textbf{Q3. How do models perform on distractors?}
As shown in Table~\ref{tab:main_results}, models exhibit substantial performance discrepancy between distractor and non-distractor questions. The box plots in Figure~\ref{fig:range} illustrate that the performance variation across models for each task type is more pronounced for distractor questions. This suggests that many models rely on linguistic priors or other heuristics rather than consistently attending to low-level acoustic features of the audio.
Figure~\ref{fig:scatter}-(b) shows a negative correlation between model performance on distractor and non-distractor questions, indicating that even high-performing models rely on non-acoustic cues rather than robust auditory processing.

\noindent\textbf{Q4. How do models respond?}
Figure~\ref{fig:qual1} and Figure~\ref{fig:qual2} illustrate LALM responses to WoW-Bench questions. We observe considerable variation in answer formats across models, with many providing direct responses without explicitly analyzing the underlying acoustic features. Notably, in the Perception task that requires vocalization identification, Gemini 2.5 Flash attempts to analyze low-level acoustic details and match them to the provided answer choices. 
In contrast, in the Cognition task involving frequency comparison between three sounds, both Gemini 2.5 Flash and Qwen2.5-Omni adopt a classification-first approach, where they initially assign each sound to a high-level category such as \textit{crisp} or \textit{pig oinking}, and infer acoustic attributes accordingly, which often leads to incorrect comparisons.
We observe similar failures in distractor-type questions (Figure~\ref{fig:qual2}). For example, when given three identical sounds, Gemini 2.5 Flash incorrectly classifies one as different and justifies its answer based on characteristics inferred from the misclassified result, despite the perceptual indistinguishability of all sounds.

For a more detailed analysis of distractor-type questions, we manually annotated a sample of 44 \textit{Remember}-type items that Gemini 2.5 Flash answered incorrectly. We categorize the errors into four types:
\begin{itemize}
  \setlength{\itemsep}{-1.5mm}
  \item \textbf{No Reasoning (NR)}: The model provides only a bare answer without explanation, e.g., \textit{Answer: B. Sound 2}.
  \item \textbf{Semantic Categorization (S)}: The model selects an answer based on high-level semantic categories rather than acoustic properties, e.g., 
    \textit{“The reference sound is a crying baby. Sound 1 is a mechanical sound. Sound 2 is a mechanical sound. Sound 3 is a crying baby. Sound 3 is the same as the reference sound. Answer: C”}.
  \item \textbf{Low-Level Grounding (L)}: The model reasons based on specific acoustic properties, e.g.,
    \textit{“The reference sound is a steady, high-pitched tone. Sound 1 is a scratching or scrubbing sound. Sound 2 is the same steady, high-pitched tone as the reference sound. Sound 3 is a low-frequency, rhythmic tapping or clicking sound. Sound 2 is the same as the reference sound. Answer: B”}.
  \item \textbf{Direct Comparison (C)}: The model engages in direct comparison with the reference sound, e.g.,
    \textit{“The audio contains a reference sound followed by three other sounds. The reference sound is a short electronic blip. Sound 1 is different from the reference sound. Sound 2 is different from the reference sound. Sound 3 is identical to the reference sound. Therefore, Sound 3 is the same as the reference sound. Answer: C”}.
\end{itemize}

\begin{table}[t]
\centering
\begin{tabular}{lccccc}
\hline
Category    & NR & S  & L & C  & \textbf{Total} \\
\hline
Count       & 16 & 16 & 2 & 10 & 44             \\
\hline
\end{tabular}%
\caption{Counts of error types in 44 incorrectly answered \textit{Remember}-type distractor questions. NR denotes No Reasoning; S denotes Semantic Categorization; L denotes Low-Level Grounding; C denotes Direct Comparison}
\label{tab:error_analysis}
\end{table}

As shown in Table~\ref{tab:error_analysis}, we observe that when models provide reasoning, semantic categorization is the most frequent strategy. This reflects a category-first behavior, as demonstrated in Figure~\ref{fig:qual2}, where the model first classifies the sounds and then infers acoustic characteristics based on those categories.
Such behavior suggests that the model is relying on heuristic shortcuts rather than genuine perceptual grounding. While two sounds may belong to the same semantic category, they can still be acoustically distinct. However, the model often treats them as identical solely based on category, failing to capture perceptual differences. This underscores the effectiveness of distractor-type questions in diagnosing whether models are truly attending to low-level acoustic detail.

These findings suggest that even when the questions require fine-grained acoustic perception, LALMs tend to prioritize semantic classification over perceptual grounding. This tendency highlights a key limitation in their ability to process and reason directly from low-level listening. Additional qualitative analyses are provided in Appendix~\ref{appen:additional_qual}.

\section{Conclusion}
We introduce WoW-Bench, a new benchmark specifically designed to rigorously evaluate the fine-grained, low-level listening, and cognitive capabilities of audio-language models using marine mammal vocalizations. 
These results highlight a critical gap in current LALM architectures: despite impressive progress in general audio understanding and instruction following, robust low-level perception remains elusive. Our findings emphasize the necessity for future research to focus on improved auditory grounding and enhanced sensitivity to acoustic detail in order to close the gap between machine and human auditory cognition. WoW-Bench provides a challenging testbed for advancing the next generation of multimodal models.

\clearpage

\section*{Limitations}
While WoW-Bench provides a rigorous and innovative framework for assessing low-level listening and cognitive processing in large audio-language models (LALMs), several limitations should be acknowledged. First, the benchmark is currently restricted to marine mammal vocalizations. Although this focus is valuable for testing out-of-distribution generalization and robustness, it represents only a narrow segment of the complex diversity found in natural acoustic environments. As a result, our findings may not directly transfer to other domains such as terrestrial bioacoustics, human speech, or complex auditory scenes with overlapping sound sources and background noise.

A further limitation concerns the task format. Our use of multiple-choice questions enables standardized comparisons and the controlled introduction of distractors, but may not fully capture the open-ended reasoning or generative abilities of modern audio-language models. There remains a gap between such discrete evaluation protocols and the continuous, often ambiguous nature of real-world auditory reasoning tasks.

We encourage future research to expand the scope of evaluation, incorporate richer and more interactive assessment paradigms, and explore cross-domain and cross-modal generalization to advance the development of truly robust and perceptually grounded audio-language models.

\section*{Ethical Considerations}
The creation and use of WoW-Bench raise several ethical considerations. All audio data in the benchmark are sourced from the publicly accessible Watkins Marine Mammal Sound Database, used with explicit permission for research purposes. Care has been taken to ensure that the dataset does not include any personally identifiable information or sensitive content. The potential deployment of audio-language models in ecological monitoring or conservation contexts must consider the ethical implications of automated decision-making, particularly regarding false positives or negatives in species identification, which could impact policy or management actions.

Understanding marine mammal vocalizations is not only of scientific interest but is also crucial for ecological monitoring and marine conservation. Marine mammals play key roles as sentinels of ocean health, and their acoustic behaviors provide unique insights into ecosystem dynamics, species distributions, and the impacts of anthropogenic activities such as shipping and climate change. However, marine bioacoustics remains an under-resourced field, with limited availability of annotated datasets, research funding, and technological tools compared to terrestrial or human-focused bioacoustics. Expanding the capabilities of machine listening through responsible audio-language models can help bridge this gap, enabling more scalable, continuous, and non-invasive monitoring of marine environments.

\section*{Acknowledgments}
This work was supported by Institute of Information \& communications Technology Planning \& Evaluation (IITP) grant funded by the Korea government (MSIT) (No.~RS-2019-II191082, SW StarLab; No.~RS-2022-II220156, Fundamental research on continual meta-learning for quality enhancement of casual videos and their 3D metaverse transformation) and  the National Research Foundation of Korea (NRF) grant funded by the Korea government (MSIT) (No.~2023R1A2C2005573).
We thank Annamaria Mesaros for her valuable feedback. We gratefully acknowledge the New Bedford Whaling Museum for granting permission to use their database for research purposes. 
Gunhee Kim is the corresponding author.

\bibliography{main}

\newpage
\appendix
\section{More Related Work}
\noindent\textbf{Bioacoustics},
the study of how animals produce and receive sound, is essential for understanding animal behavior and monitoring ecosystems \cite{bioacoustic_survey}. From a machine learning perspective, it offers a rich testbed for evaluating auditory perception due to its diverse acoustic environments, wide frequency ranges, and overlapping vocalizations \cite{birdset}. Leveraging these properties, BirdSet \cite{birdset} introduces a large-scale benchmark for avian vocalizations, highlighting the potential of bioacoustic data for evaluating model robustness and distributional generalization in audio classification tasks.
We extend this line of work by focusing on marine mammal vocalizations, which span an exceptionally broad frequency range, from below 20 Hz to over 20 kHz, even up to 100 kHz~\cite{google_whale}, and are underrepresented in standard datasets, even compared to bird sounds. For example, VGGSound~\cite{vggsound} includes over 10 species-specific labels for birds but only 2 coarse labels for marine mammals. These characteristics make whale sounds a challenging domain for robust evaluation of low-level listening and out-of-distribution generalization in LALMs. Our work differs from the domain adaptation of LALMs to the bioacoustic domain \cite{naturelm_audio} in that we aim to assess the detailed perceptual capabilities of LALMs in novel auditory environments, rather than focusing on the bioacoustics task itself.

\section{Additional Experimental Results}

\begin{figure}[t]
\centering
\includegraphics[width=\linewidth]{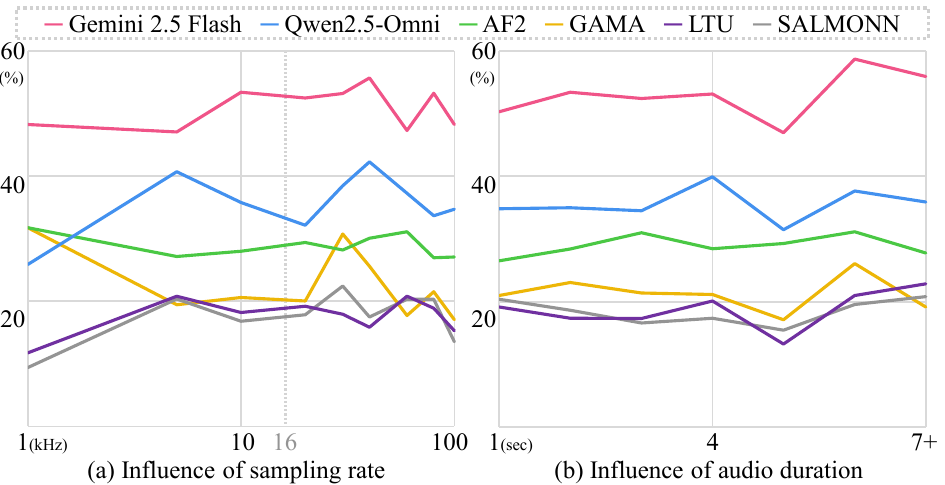}
\caption{
Influence of audio property on accuracy, namely (a) sampling rate and (b) duration.
}
\label{fig:trendline}
\vspace{-10pt}
\end{figure}

\subsection{Does acoustic range affect the performance?}
By utilizing the wide range of sample rates and durations of samples in WoW-Bench, we analyze how model performance varies with these acoustic properties. As shown in Figure~\ref{fig:trendline}, \textit{performance fluctuates with both duration and sample rate, but no sharp increase or drop is observed at any specific threshold.} This suggests that the benchmark fairly reflects the capabilities of open-source models, which typically operate on 16~kHz audio.
We hypothesize that this phenomenon arises from two factors: (1) while a considerable number of vocalizations extend beyond 10~kHz (\textit{e.g.}, dolphin whistles), a substantial portion of others have their key components within the audible range that most models can process, regardless of the original sample rate of the recordings; and (2) current models do not exhibit significantly improved low-level listening performance even within their nominal frequency range.

These findings suggest two directions for future improvement: (1) enhancing perceptual fidelity within the standard listening range to make better use of accessible acoustic information, and (2) expanding model capacity to process a broader acoustic spectrum, thereby improving generalization to a wider range of real-world audio events.

\begin{table*}[ht]
\centering
\resizebox{0.9\textwidth}{!}{
\begin{tabular}{l|ccccc}
\hline
Model Name & Remember & Understand & Apply-Frequency & Apply-Duration & Average \\
\hline
Audio Flamingo 2 (WoW-Bench)              & 5.0  & 30.5 & 32.5 & 29.5 & 24.4 \\
Audio Flamingo 2 (ESC-50)                 & 24.0 & 48.0 & 32.0 & 28.0 & 33.0 \\
\cdashline{0-5}
Qwen2.5-Omni (WoW-Bench)     & 21.5 & 46.3 & 34.5 & 42.0 & 36.1 \\
Qwen2.5-Omni (ESC-50)        & 20.0 & 76.0 & 53.0 & 40.0 & 47.3 \\
\cdashline{0-5}
Gemini 1.5 Pro (WoW-Bench)   & 34.5 & 33.7 & 32.5 & 37.5 & 34.6 \\
Gemini 1.5 Pro (ESC-50)      & 44.0 & 80.0 & 44.0 & 44.0 & 53.0 \\
\cdashline{0-5}
Gemini 2.0 Flash (WoW-Bench) & 54.0 & 48.4 & 35.5 & 14.5 & 38.1 \\
Gemini 2.0 Flash (ESC-50)    & 68.0 & 76.0 & 28.0 & 24.0 & 49.0 \\
\hline
\end{tabular}
}
\caption{Evaluation results on non-distractor Cognition questions from WoW-Bench and ESC-50 based set.}
\label{tab:esc_50}
\end{table*}

\subsection{How do models perform on in-domain sounds?}\label{appen:esc50}
We analyze how the use of OOD audio impacts performance on the Cognition tasks, as the low-level listening tasks can also be constructed using in-domain audio. We generate 100 questions based on sounds from the ESC-50 \cite{esc50} dataset that correspond to four Cognition tasks: \textit{Remember}, \textit{Understand}, \textit{Apply-Frequency}, and \textit{Apply-Duration}. The overall performance on Non-Distractor Cognition tasks is summarized in Table~\ref{tab:esc_50}. The model performance is generally higher on the in-domain questions than on WoW-Bench, which implies that the challenging nature of our benchmark largely stems from the perception of OOD sounds rather than the tasks themselves. We observe that the improvement is especially pronounced for certain tasks like \textit{Understand}. Furthermore, Qwen2.5-Omni, which adopts the classification-first strategy in the \textit{Apply-Frequency} task as shown in Figure 4-(b), achieved significant score improvements of nearly 20\% on this task.

In addition, while models generally improved on the in-domain \textit{Remember} and \textit{Understand} tasks, the patterns were mixed in the \textit{Apply} tasks. This indicates that some improvements may stem from models leveraging prior knowledge of familiar sources, whereas other tasks still expose their limited listening ability. This mixed tendency implies that it is difficult to disentangle whether models know the sound or listen to it using familiar sound domains.
To provide a more controlled perspective, we believe OOD inputs are crucial in evaluating low-level listening, as they reduce the influence of acoustic knowledge about the classified sound event. This motivation underlies our design of the Cognition tasks and is further supported by our human evaluation, which shows that humans can solve these tasks effectively without prior familiarity with the domain by relying primarily on listening.

\subsection{Additional Qualitative Analysis}\label{appen:additional_qual}

We conducted a supplementary error-type analysis to provide more diagnostic insight. Specifically, we examined how the presence of certain words in Gemini 2.5 Flash’s responses correlates with answer correctness.

\begin{table*}[t]
\centering
\resizebox{0.9\textwidth}{!}{%
\begin{tabular}{lcc lcc lcc}
\hline
\multicolumn{3}{c}{(a) Perception} & \multicolumn{3}{c}{(b) Non-distractor} & \multicolumn{3}{c}{(c) Distractor} \\
\cmidrule(r){1-3} \cmidrule(r){4-6} \cmidrule(r){7-9}
Word         & Fraction & Accuracy & Word        & Fraction & Accuracy & Word        & Fraction & Accuracy \\
\midrule
hum          & 25/40    & 0.6250   & longest     & 152/182  & 0.8352   & wide        & 29/52    & 0.5577   \\
chirp        & 29/54    & 0.5370   & reference   & 103/134  & 0.7687   & across      & 20/40    & 0.5000   \\
modulated    & 19/36    & 0.5278   & duration    & 186/265  & 0.7019   & consistent  & 12/29    & 0.4138   \\
present      & 26/55    & 0.4727   & silence     & 189/278  & 0.6799   & present     & 11/32    & 0.3438   \\
continuous   & 35/75    & 0.4667   & end         & 91/146   & 0.6233   & modulating  & 11/32    & 0.3438   \\
\textbf{Overall}    &          & 0.3716   &             &          & 0.5453   &             &          & 0.1833   \\
\hline
\multicolumn{3}{c}{(a) Perception} & \multicolumn{3}{c}{(b) Non-distractor} & \multicolumn{3}{c}{(c) Distractor} \\
\cmidrule(r){1-3} \cmidrule(r){4-6} \cmidrule(r){7-9}
Word         & Fraction & Accuracy & Word        & Fraction & Accuracy & Word        & Fraction & Accuracy \\
\hline
frequency    & 14/52    & 0.2692   & broadband   & 119/334  & 0.3563   & lower       & 3/35     & 0.0857   \\
dolphin      & 35/131   & 0.2672   & harmonic    & 43/123   & 0.3496   & occur       & 5/62     & 0.0806   \\
sperm        & 14/53    & 0.2642   & repetitive  & 48/140   & 0.3429   & third       & 3/61     & 0.0492   \\
specie       & 23/103   & 0.2233   & pulse       & 50/152   & 0.3289   & comparing   & 4/115    & 0.0348   \\
common       & 9/48     & 0.1875   & modulating  & 45/140   & 0.3214   & longest     & 1/42     & 0.0238   \\
\textbf{Overall}    &          & 0.3716   &             &          & 0.5453   &             &          & 0.1833   \\
\hline
\end{tabular}%
}
\caption{Keyword-based accuracy for the Gemini 2.5 Flash model.
The top group displays exemplar keywords with high accuracies, while the bottom group displays keywords with low accuracies. 
Each cell shows the fraction of correct responses when the specified keyword appears in the model’s answer, broken down by question type: (a) Perception tasks, (b) Cognition tasks without distractors, and (c) Cognition tasks with distractors.}
\label{tab:keyword_analysis}
\end{table*}

The results are shown in Table~\ref{tab:keyword_analysis}. For \textit{Perception}-type questions, Gemini 2.5 Flash performs relatively well when the response includes acoustic event descriptions such as “hum” or “chirp.” However, the model often struggles with identifying specific species, particularly dolphin vocalizations.
For \textit{Cognition} questions without distractors, the model performs better when its responses include temporal or presence-related terms such as “longest,” “silence,” or “occurs.” Interestingly, in contrast to Perception tasks, performance degrades when responses include acoustic feature terms such as “repetitive”.
For \textit{Distractor}-type questions, the trend shifts again; overall performance is lower, but the model shows slightly improved alignment when using global contrastive terms such as “across,” “consistent,” or “wide.” However, it tends to fail in relative comparison, e.g., terms like “longest,” “lower,” or “third,” which it handled relatively well in non-distractor Cognition tasks.

These findings suggest that the reasoning chains used in Cognition tasks differ meaningfully from those used in Perception tasks, and that such reasoning fails to transfer to distractor conditions. This supports our claim that distractor-type questions pose a meaningful challenge and reveal limitations in models’ generalization of reasoning strategies. The error-type patterns we observe here are consistent with trends seen in our qualitative analysis 

\subsection{Evaluation of Speech-based LALMs}

\begin{table*}[ht]
\centering
\resizebox{0.9\textwidth}{!}{%
\begin{tabular}{l|cccc|ccccccccc|c}
\hline
\multirow{2}{*}{\textbf{Model}} 
  & \multicolumn{4}{c|}{\textbf{Perception}} 
  & \multicolumn{9}{c|}{\textbf{Cognition}} 
  & \multirow{2}{*}{\textbf{Total}} \\
  & \textbf{S} & \textbf{V} & \textbf{B} & \textbf{All} 
  & \textbf{R} & \textbf{U} & \textbf{AF} & \textbf{AD} & \textbf{AA} & \textbf{AV} & \textbf{ND} & \textbf{D} & \textbf{All} \\ 
\hline
Gemini 2.5 Flash & 27.3 & 54.6 & 30.0 & 37.2 
                 & 64.4 & 39.2 & 36.0 & 66.4 & 34.0 & 42.7 & 54.5 & 18.3 & 47.2 
                 & 45.5 \\
GPT-4o-Audio     & 19.2 & 38.1 & 19.0 & 25.3 
                 & 14.0 & 26.3 & 21.2 & 16.0 & 24.8 & 25.3 & 23.0 & 14.0 & 21.2 
                 & 21.9 \\
\hline
\end{tabular}%
}
\caption{Evaluation result of Gemini 2.5 Flash and GPT-4o-Audio on the WoW-Bench.}
\label{tab:gpt}
\end{table*}

Our goal was to provide a broad and representative evaluation of current LALMs, with a focus on models capable of processing general, non-speech sound events. We intentionally excluded speech-only language models, since our tasks target non-speech auditory processing.

We additionally evaluated GPT-4o-Audio \cite{gpt4o_audio}, but as shown in Table~\ref{tab:gpt}, its performance was substantially lower than that of Gemini 2.5 Flash across both the Perception and Cognition benchmarks. In several cases, GPT-4o-Audio produced fallback responses such as “My capabilities include text-based information and analysis, but not audio-based identification,” indicating limited ability to handle audio input. These observations support our decision to focus on models explicitly designed for general audio processing, rather than those optimized primarily for speech.

\section{Details of WoW-Bench}
\begin{table}[ht]
\centering
\resizebox{\linewidth}{!}{
\begin{tabular}{llccc}
\hline
Task & Subtask & ND & D & Total \\
\hline
\multirow{3}{*}{Perception} & Species & 99  & - & 99  \\
                            & Vocalization & 97 & -  & 97 \\ & Both & 100 & - & 100 \\
\hline
\multirow{6}{*}{Cognition}  & Remember & 200 & 50 & 250 \\
                            & Understand & 190 & 50 & 240 \\
                            & Apply-Frequency & 200  & 50 & 250 \\
                            & Apply-Duration & 200  & 50 & 250\\
                            & Analyze-Acoustics & 200 & 50 & 250 \\
                            & Analyze-Vocalization & 191 & 50 & 241 \\
\hline
\multicolumn{2}{c}{Total}   & 1477 & 300 & 1777  \\
\hline
\end{tabular}
}
\caption{Number of questions in WoW-Bench by task type. ND denotes non-distractor questions, and D denotes distractor questions.}
\label{tab:statistics}
\end{table}

\subsection{Statistics of WoW-Bench}
Our WoW-Bench consists of 1,777 rigorously validated question-answer pairs with audios.
Perception benchmark consists of 296 pairs, with 99, 97, and 100 pairs for Species, Vocalization, and Both task, respectively.
Each of the six tasks for non-distractor cognition benchmark contains 200 pairs except for Understand and Analyze-Vocalization, for which we filtered out 10 and 9 pairs for their quality.
Lastly, regarding distractor-based cognition, we secure a total of 300 pairs, \textit{i.e.}, 50 for each task type. The number of questions for each task type is detailed in Table~\ref{tab:statistics}.

\subsection{Distractor Design Process}
\label{appen:distractor_design}
We systematically design the distractors by inverting the expected pattern for each question type, thereby encouraging the model to rely on listening to solve the task.
For questions that require comparison across multiple audio samples, we invert the expectation that one of the labeled candidate sounds corresponds to the correct answer. In Remember, all options differ from the reference, making “None of the above” the correct answer. In Apply, all provided sounds are acoustically identical, and the model must correctly select “All are indistinguishable.” For tasks focused on understanding the acoustic characteristics of input, we reverse the expectation that the input contains meaningful acoustic content. In Understand, the reference audio is replaced with synthetic noise, requiring the model to avoid hallucinating a semantic interpretation. In Analyze–Acoustics, one segment of a sequence is replaced with noise, and the model must identify the disrupted transition. Lastly, in Analyze–Vocalization, which typically involves detecting species transitions, we eliminate such transitions by concatenating two identical or same-species vocalizations. The model must detect structural redundancy or similarity, rather than blindly assuming a cross-species transition.

To ensure consistency of answer choices across distractor and non-distractor questions, we include the distractor-style options (\textit{e.g.}, “None match” or “All indistinguishable”) in Cognition questions. We generate distractor variants by selecting a subset of non-distractor questions and replacing their audio components with distractor audio. For noise-based distractors, we sample from a diverse set of synthetic noise types (\textit{e.g.}, white, pink, brown, blue) that matches the duration and sampling rate of randomly selected real audio clips to ensure consistency in acoustic conditions.

\subsection{Details of Data Curation}
\label{appen:data_curation}
With the exception of Remember questions, which are manually constructed by selecting candidate sounds, all other questions are generated using a large language model. We provide the model with relevant background information and metadata of the audio clips as input. For questions that require acoustic details not available in the metadata (\textit{i.e.}, Understand and Apply), we employ a vision language model and additionally provide spectrograms of the audio clips. For Analyze-Acoustics, we reuse and adapt acoustic descriptions previously generated and validated from the Understand task to construct transition-based questions.  
For all question types, we use GPT-4o (\texttt{gpt-4o-2024-11-20}) as both the large language model and the vision language model.

For the Perception benchmark, we use metadata such as species names and vocalization descriptions to generate questions and answer choices using GPT-4o~\cite{gpt4o}. For the Cognition benchmark, Remember questions are constructed by manually selecting candidate sounds. Since the Understand and Apply tasks require fine-grained descriptions of acoustic features that cannot be derived from metadata alone, we provide spectrograms to GPT-4o along with detailed guidance for spectrogram interpretation to generate appropriate questions and answer choices.
For Analyze–Acoustics, we reuse and adapt acoustic descriptions previously generated and validated from the Understand task to construct transition-based questions. For Analyze–Vocalization, metadata is used to create plausible transitions between species and vocalization types. In both Analyze tasks, GPT-4o is used to generate the final question and answer sets based on selected audio clips. 

For each Perception task, we generate 100 questions. For the Cognition tasks, including the subtasks within Apply and Analyze, we generate 200 questions along with 50 adversarial distractor questions per task.
We place greater emphasis on the Cognition tasks, as reflected by the higher number of Cognition questions, for two reasons: (1) the Perception questions partly rely on prior knowledge on marine mammals, which current LALMs may not be well-equipped to handle, and (2) the Cognition tasks are more solvable without such prior knowledge, relying primarily on low-level listening. As a result, we treat Cognition scores as a more reliable indicator of perceptual ability in isolation when reporting overall benchmark performance.

All generated questions undergo rigorous human verification and reannotation. Each generated question-answer pair is classified into one of three categories: (1) accept as is, (2) accept with revision, or (3) discard. Similar to the validation pipeline used in MMAU \cite{mmau}, each question is reviewed by three experienced annotators who cross-check the prompts and QA pairs against the associated metadata and audio clips. If the pair requires revision, annotators evaluate whether it can be rectified with minor edits, such as factual corrections or refinement of uninformative answer choices. If the issue is not fixable, for example, due to high ambiguity or incorrect source data, the pair is discarded. Among approximately 1,900 generated questions, 112 were discarded and 485 were revised, where revisions ranged from minor edits to ensure text style consistency to updates of factual content. During this process, answer choices are also paraphrased to prevent models from exploiting surface-level lexical cues.

A portion of WoW-Bench was included in the held-out test set of 
the Bioacoustics QA track of the DCASE 2025 Audio Question Answering 
Challenge~\citep{mdaudio2025}. The Bioacoustics QA training and development sets consist of disjoint question-answer pairs, although all audio is drawn 
from the same source, the Watkins Marine Mammal Sound Database~\citep{watkins}. 
We note this relationship for transparency.

\begin{figure*}[t]
\centering
\includegraphics[width=\textwidth]{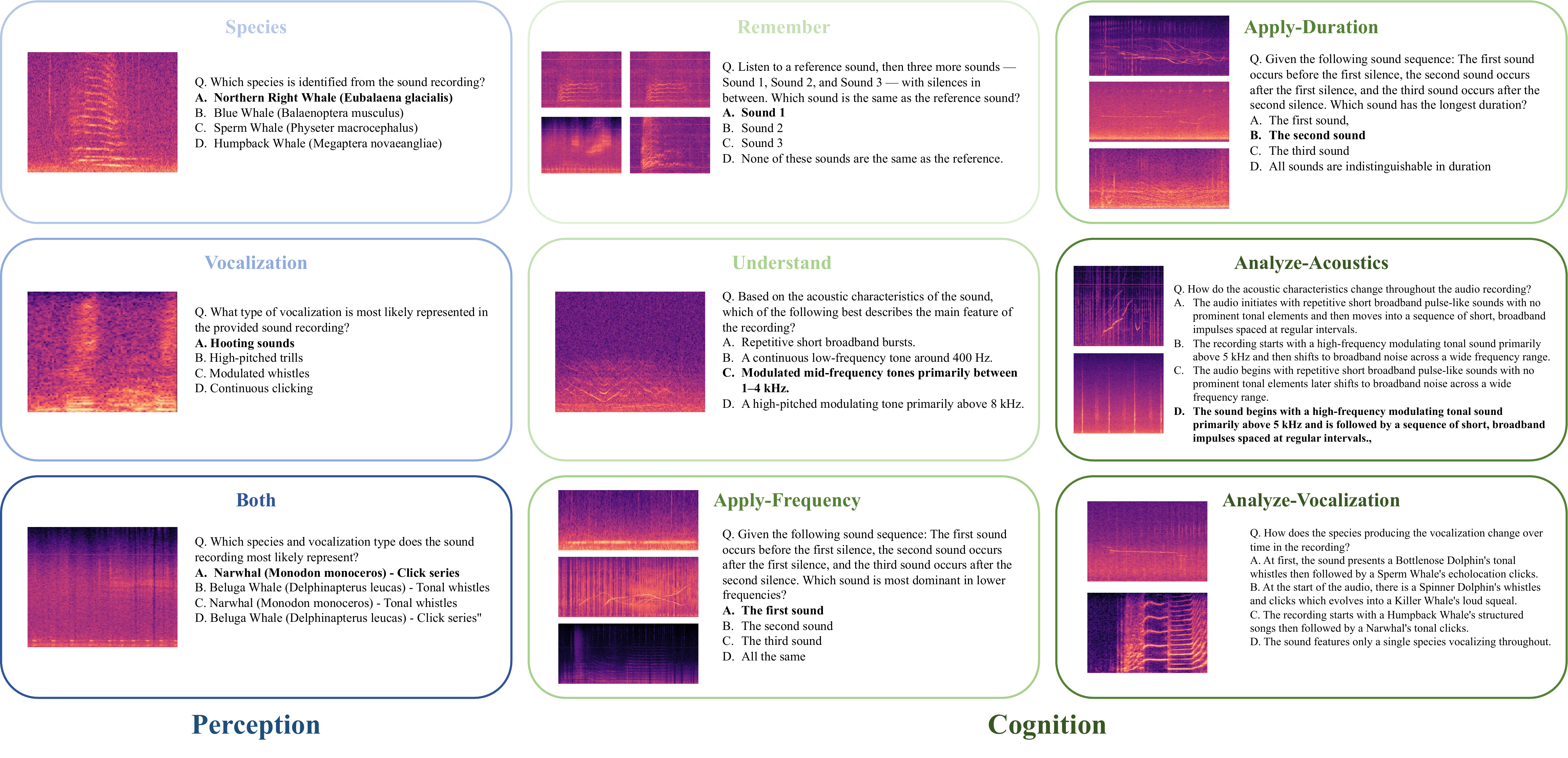}
\caption{Example questions from each task type in WoW-Bench, shown alongside spectrograms of the corresponding input audio.}
\label{fig:sample_questions}
\vspace{-10pt}
\end{figure*}
\subsection{Sample Questions of WoW-Bench}
Figure~\ref{fig:sample_questions} presents representative examples of WoW-Bench questions across all types of tasks.

\subsection{Analysis on the answer choice}
\begin{table}[h!]
\centering
\resizebox{\linewidth}{!}{
\begin{tabular}{lcccc}
\hline
Model & Original & Shuffle 1 & Shuffle 2 & STD \\
\hline
Audio Flamingo 2    & 31.9 & 30.7 & 31.0 & 0.666 \\
Qwen2-Audio-Instruct & 32.7 & 33.4 & 32.1 & 0.636 \\
Gemini 2.0 Flash & 35.0 & 36.8 & 36.3 & 0.958 \\
\hline
\end{tabular}}
\caption{Model performance across different shuffle configurations.}
\label{tab:choice_bias}
\end{table}
To mitigate potential choice bias, we randomize the order of answer options during benchmark construction. This does not apply to the Remember and Apply questions, where the sequential presentation of options serves a functional purpose (e.g., “A. Sound 1, B. Sound 2, C. Sound 3”). For all other question types, we randomize the choice order and matched a uniform distribution to minimize bias (A: 25.9\%, B: 23.0\%, C: 27.6\%, D: 23.5\%).

In addition, we examine the effect of choice shuffling by conducting two additional evaluations. For all 1,072 questions outside the Remember and Apply types, we shuffle the answer options and evaluate the model performance on these randomized datasets. The results, summarized in Table~\ref{tab:choice_bias}, show that the standard deviation of accuracy across shuffles is less than 1\%, indicating that the models produce consistent results regardless of the choice order. This suggests that the randomization strategy effectively mitigates choice bias in WoW-Bench.

\subsection{Clarification on Design Choices}
\noindent\textbf{Q. Why only marine mammal vocalizations?}
We intentionally designed the benchmark using marine mammal vocalizations to construct a meaningful out-of-distribution (OOD) setup. LALMs often bypass low-level acoustic perception by inferring answers based on high-level semantic associations with known sounds. This behavior undermines the evaluation of genuine low-level listening capabilities, as shown in qualitative examples in Figure~\ref{fig:qual1}, and Figure~\ref{fig:qual2} . 

Moreover, identifying suitable OOD audio content is non-trivial, as the pre-training data of many LALMs already cover a broad range of everyday sounds, including human speech and urban noise. Even when we modestly extended the dataset by adding other animal vocalizations from the VGGSound test set and asked the model to classify the animal species, performance significantly improved. For instance, Qwen2-Audio-Instruct’s accuracy on the species classification task increased from 28.3\% in WoW-Bench to 76\% in the VGGSound-based set (e.g., Baltimore oriole, pigeon, parrot). This result suggests that broadening the sound categories may compromise the OOD nature of the task and reduce the benchmark’s ability to isolate low-level acoustic reasoning. We believe that its focused and controlled design using marine mammal vocalizations enables a more precise evaluation of models’ low-level listening capabilities in genuinely OOD settings.

\noindent\textbf{Q. Why use only four levels from Bloom’s taxonomy?} Bloom’s taxonomy consists of six levels of cognitive objectives: Remember, Understand, Apply, Analyze, Evaluate, and Create. Among these, Evaluate and Create are less suitable for modeling low-level listening assessment, since they involve subjective judgment, which are difficult to score objectively, or creative synthesis of new audio. These levels go beyond the scope of low-level listening and the current capabilities of LALMs, which generally cannot generate non-speech audio. Therefore, we focus our benchmark on the first four levels, which are better aligned with the tasks requiring grounded, objective responses based on low-level listening.

\noindent\textbf{Q. Why multiple-choice question format?} First, multiple-choice question (MCQ) format is widely adopted in prior LALM benchmarks~\cite{mmau, airbench} due to its scalability, objective evaluation, and ease of reproducibility. It enables fine-grained control over task structure and difficulty, and facilitates direct comparison across models without requiring prompt-specific evaluation metrics or human scoring.
Second, the generative description of marine mammal vocalizations poses a unique challenge. These sounds are difficult to describe even for human annotators, and current models often default to familiar semantic categories instead of grounding their responses in low-level acoustic properties. For example, when prompted in a manner similar to the Understand task to describe a marine mammal vocalization, the model responded as follows

\begin{quote}
\textbf{Gemini 2.5 Flash:} Based on the audio provided from 00:00 to 00:04, the acoustic characteristics are:
\begin{itemize}
    \item \textbf{Source:} Avian Vocalizations (specifically bird calls).
    \item \textbf{Type:} The calls are resonant, distinct, and have a somewhat squawking or crying quality, rather than being simple chirps or melodic songs.
    \item \textbf{Pattern:} There seems to be a noticeable rhythmic or patterned quality to the sequence of calls.
    \item \textbf{Number:} It sounds like multiple birds are vocalizing.
    \item \textbf{Frequency:} The sound primarily occupies the mid-to-high frequency range, typical of bird calls, but with a resonant lower component in the calls.
    \item \textbf{Dynamics:} Within this short clip, the sound is relatively consistent in volume.
\end{itemize}
\end{quote}

\begin{quote}
\noindent\textbf{Reference description:} A series of modulated mid-to-high pitched tones primarily between 2–8~kHz.
\end{quote}

\medskip

The model’s output combines high-level semantic content (bird call) with limited references to low-level acoustic features (mid-to-high frequency). However, in a free-form format, such responses are difficult to evaluate reliably, since they vary in specificity and may omit key acoustic details necessary for assessing perceptual grounding.

Third, while our benchmark adopts an MCQ format, it does not preclude open-ended reasoning. Models such as Qwen2.5-Omni and Gemini 2.5 Flash demonstrate clear intermediate reasoning steps before selecting a final answer as shown in Figure~\ref{fig:qual1}, the MCQ structure simply constrains the output space for evaluation purposes, while still allowing models to engage in internal multi-step reasoning, inference, and comparison.

\noindent\textbf{Q. Why evaluate species task in a zero-shot setting?} 
The goal of the species task is to assess whether models can leverage their knowledge about whale species and their vocalizations, potentially acquired from text sources such as Wikipedia, in combination with low-level acoustic cues to make informed predictions.
In our results, both models and human participants perform near random chance, which confirms that the task domain effectively represents an OOD scenario. This OOD nature, along with the task’s design, highlights a valuable direction for future work: explicitly bridging low-level auditory perception with external knowledge to address the performance gap in such challenging scenarios.

\section{Experimental Details}
\label{app:sec:exp}

\subsection{Models}
We enumerate the models and implementations used in all of our experiments as follows:
\begin{itemize}
   \item \textbf{SALMONN}~\cite{salmonn}\footnote{\url{https://github.com/bytedance/SALMONN}} (Apache-2.0)
   \item \textbf{LTU}~\cite{ltu}\footnote{\url{https://github.com/YuanGongND/ltu}} (CC BY Attribution 4.0 International)
   \item \textbf{LTU-AS}~\cite{ltu_as}\footnote{\url{https://github.com/YuanGongND/ltu}} (CC BY Attribution 4.0 International)
   \item \textbf{GAMA}~\cite{gama}\footnote{\url{https://github.com/Sreyan88/GAMA}} (Apache-2.0)
   \item \textbf{Qwen-Audio-Chat}~\cite{qwen_audio_1}\footnote{\url{https://github.com/QwenLM/Qwen-Audio}} (Tongyi Qianwen LICENSE AGREEMENT)
   \item \textbf{Qwen2.5-Omni-7B}~\cite{xu2025qwenomni}\footnote{\url{https://github.com/QwenLM/Qwen2-Audio}}  (Apache-2.0)
   \item \textbf{Audio Flamingo 2}~\cite{af2}\footnote{\url{https://github.com/NVIDIA/audio-flamingo/tree/audio_flamingo_2}} (MIT License)
   \item \textbf{Audio Flamingo 3}~\cite{af3}\footnote{\url{https://github.com/NVIDIA/audio-flamingo/tree/audio_flamingo_3}} (MIT License)
   \item \textbf{Gemini 1.5 Pro}~\cite{gemini1.5}\footnote{\url{https://arxiv.org/abs/2403.05530}}
   \item \textbf{Gemini 2.0 Flash}~\cite{gemini}\footnote{\url{https://cloud.google.com/vertex-ai/generative-ai/docs/models/gemini/2-0-flash}}
   \item \textbf{Gemini 2.5 Flash}~\cite{gemini2.5}\footnote{\url{https://cloud.google.com/vertex-ai/generative-ai/docs/models/gemini/2-5-flash}}
   \item \textbf{Gemini 2.5 Pro}~\cite{gemini2.5}\footnote{\url{https://cloud.google.com/vertex-ai/generative-ai/docs/models/gemini/2-5-pro}}
   \item \textbf{GPT-4o-Audio}~\cite{gpt4o_audio}\footnote{\url{https://platform.openai.com/docs/models/gpt-4o-audio-preview}}
\end{itemize}

To the best of our knowledge, we confirm that our use of aforementioned scientific artifacts is fully compliant of their intended use.
All reported results are based on a single run per experiment, where we use one NVIDIA RTX A6000 and 8 CPU cores for running inference with open models.

\subsection{Prompts}
To ensure fair and standardized evaluation across diverse audio-language models (LALMs) we design task-specific prompts for all inference tasks in WoW-Bench. For multiple-choice questions, the prompt structure clearly presents the question, audio context, and the set of answer options, followed by explicit instructions for answer selection. For example, in the Perception and Cognition benchmarks, a typical prompt is:

\begin{quote}
\textit{You will listen to a series of audio recordings. Based on what you hear, choose the most appropriate answer from the options below. Reply with the letter corresponding to your choice.}
\end{quote}

Depending on the capabilities and response tendencies of each model, we experiment with minor variations such as: “Please select the correct answer,” “Return only the answer letter,” or “Explain your reasoning, then provide the answer letter.” The best-performing prompt for each model is selected based on preliminary validation.

\subsection{Automated Answer Extraction and GPT Evaluation}\label{appen:eval_detail}

Given the diverse response formats of LALMs, we employ an automated extraction pipeline to reliably determine model answers. Model outputs are parsed using a lightweight regular expression matcher to identify the final answer letter, regardless of the presence of additional text or reasoning.

For evaluation and quality assurance, we use GPT-4.1-mini to resolve ambiguous cases where the model output does not clearly map to a single answer choice. The evaluation prompt is as follows:

\begin{quote}
\textit{Given the following question, options, and the model’s response, identify the answer letter (A, B, C, or D) selected by the model. If no clear answer is provided, return “Invalid.”}
\end{quote}

This automated approach ensures consistent and scalable evaluation, particularly for open-ended or verbose outputs. Ambiguous or invalid responses are excluded from accuracy calculations.

\begin{table*}[t]
\centering
\resizebox{0.9\textwidth}{!}{
\begin{tabular}{lcccc}
\toprule
Category                 & Gemini 2.5 Flash (\%) & Non-Expert (\%) & Audio Expert (\%) & Human Total (\%) \\
\midrule
Perception               & 41.67                 & \textbf{46.67}           & 33.33             & 40.00            \\
Cognition: Non-distractor & 48.33                 & 62.67           & \textbf{78.67}             & 70.67            \\
Cognition: Distractor    & 25.00                 & 65.00           & \textbf{88.33}           & 76.67            \\
Remember                 & 57.14                 & \textbf{97.14}  & \textbf{97.14}             & \textbf{97.14}            \\
Understand               & 42.86                 & 57.14           & \textbf{77.14}             & 67.14            \\
Apply–Frequency          & 21.43                 & 65.71           & \textbf{85.71}            & 75.71            \\
Apply–Duration           & 57.14                 & 80.00           & \textbf{94.29}            & 87.14            \\
Analyze–Acoustics        & 21.43                 & 42.86           & \textbf{77.14}             & 60.00            \\
Analyze–Vocalization     & 50.00                 & 37.14           & \textbf{57.14}            & 47.14            \\
\bottomrule
\end{tabular}%
}

\caption{Detailed human evaluation results across the various task categories.}
\label{tab:human_eval}
\end{table*}

\subsection{Human Evaluation}\label{appen:human_eval}
To quantitatively assess human performance on the WoW-Bench tasks, we conducted a targeted human evaluation on a stratified subset of our dataset. From the full benchmark, we select a subset that is representative of the original dataset. To achieve this, we first randomly sampled five mini test sets, each composed of 108 question-answer pairs. We then evaluated the performance of three models (Audio Flamingo 2, Qwen2-Audio, and Gemini 2.5 Flash) on each of them. For every model and subset, we computed the performance difference between the mini test set and the full dataset, and then averaged these differences across the models. We selected the subset with the smallest average difference from the full dataset.
This approach ensured that the selected subset was both representative and appropriate for reliable human-model comparison.

\begin{figure*}
    \centering
    \includegraphics[width=\textwidth]{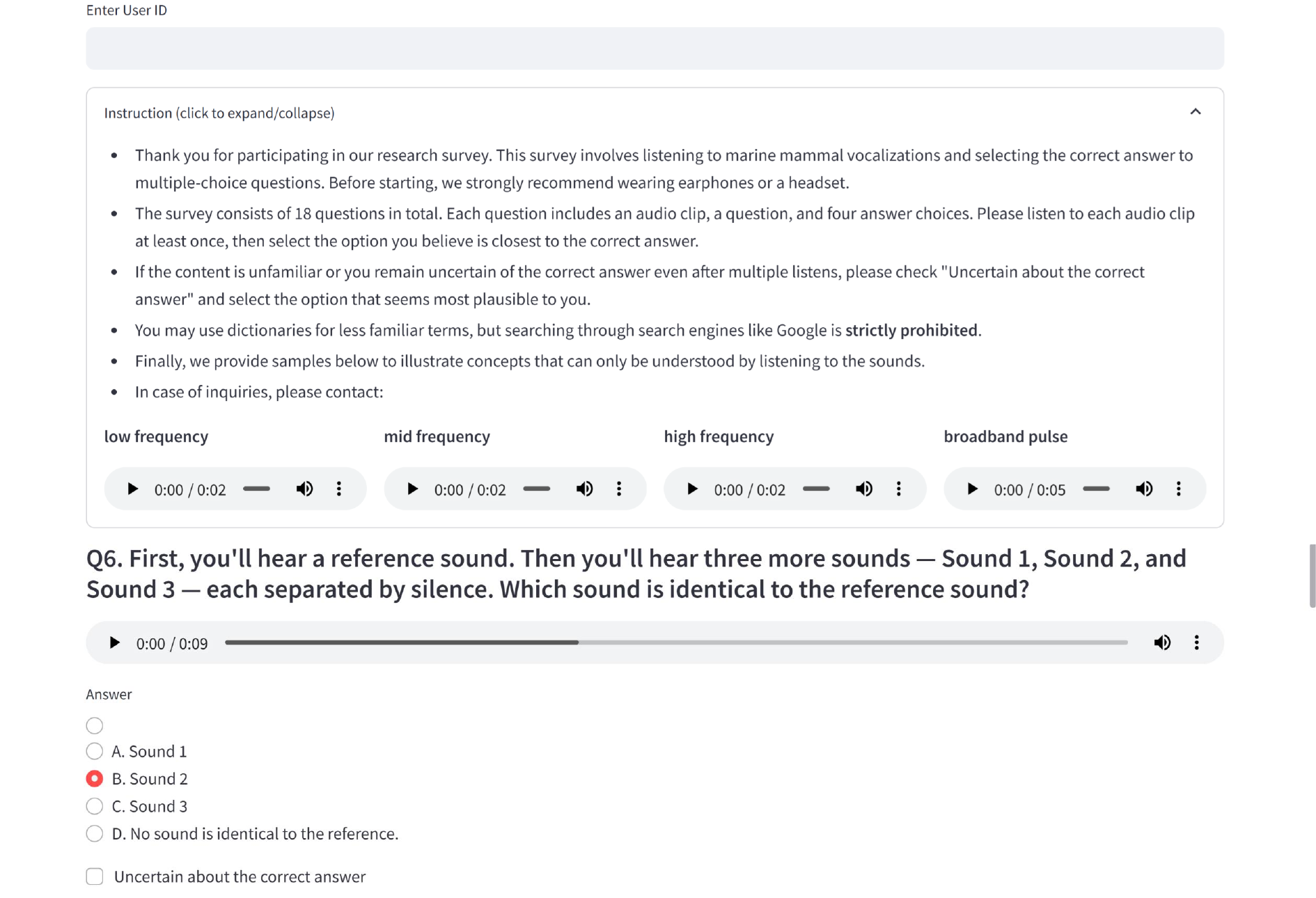}
    \caption{Multiple-choice questionnaire interface used for human evaluation.}
    \label{fig:app_interface}
\end{figure*}

The 108 questions were divided into two questionnaires, each comprising 54 items, as illustrated in Figure~\ref{fig:app_interface}. Each questionnaire was independently completed by five participants, resulting in a total of ten unique annotators and 5 times redundancy for every question. In detail, we recruited five inexperienced listeners and five participants with audio-related expertise (e.g., individuals with multiple publications in relevant areas). Detailed results across tasks and participant types are provided in Table~\ref{tab:human_eval}. 

Participants were recruited on a voluntary basis and provided informed consent, with explicit communication regarding the use and anonymization of their response data. Annotators completed the survey remotely, in their environment of choice, and were free to pause or discontinue at any time to mitigate fatigue effects. The user interface was designed for clarity and accessibility: each question included audio playback controls, clearly labeled answer choices, and an optional “Uncertain about the correct answer” option to capture genuine uncertainty and discourage forced guessing.

Detailed instructions were provided at the start of the survey, including recommendations for headphone use and prohibitions on the use of search engines or external reference materials. To minimize potential bias, illustrative audio examples were provided for technical terms such as ``\textit{low frequency},'' ``\textit{high frequency},'' and ``\textit{broadband pulse},'' allowing participants to anchor their perceptual judgments to auditory references rather than textual definitions.

Participants were instructed to base their responses solely on the presented audio and to avoid making value-based or speculative judgments. Average completion time was less than one hour per participant, and all participants received compensation in accordance with the local legal minimum wage to ensure ethical standards for research participation. This proposed protocol was designed to balance experimental rigor with participant well-being, producing high-quality human baseline data for comparison with model predictions.

To enable statistical assessment of the human responses, we measured inter-rater agreement using Fleiss' kappa~\cite{fleiss1971kappa}, with the results summarized in Table~\ref{tab:fleiss_kappa}.
Low kappa values imply an absence of systematic bias in the answer selection. However, we observed a relatively higher agreement pattern in the Cognition tasks, where the kappa score was approximately three times higher than in the Perception tasks. In contrast, the low kappa in the Perception benchmark, combined with accuracy below the chance ($\bar{P}_e$) level, indicates that the choices were nearly equivalent to blind guessing. This also relates to the counterintuitive ranking among Gemini 2.5 Flash, non-experts, and experts for the Perception benchmark.

Regarding non-expert performance relative to Gemini 2.5 Flash, the Cognition tasks primarily target low-level listening skills that most humans possess inherently. Many tasks in this category do not require specialized acoustic knowledge, e.g., recognizing identical sounds in Remember or comparing audio lengths in Apply-Duration. It is therefore unsurprising that non-experts can outperform state-of-the-art LALMs on these tasks. Notably, in the tasks that require domain-specific knowledge, such as Analyze-Vocalization, Gemini 2.5 Flash surpasses non-expert performance.

\begin{table}[h]
\centering
\resizebox{\linewidth}{!}{\begin{tabular}{lccc}
\hline
 & Accuracy & Chance ($\bar{P}_e$) & kappa ($\kappa$) \\
\hline
Perception & 0.4444 & 0.5062 & 0.0775 \\
Cognition  & 0.7421 & 0.6172 & 0.2329 \\
Total      & 0.6759 & 0.5619 & 0.2476 \\
\hline
\end{tabular}}
\caption{Inter-rater agreement measured by Fleiss' kappa.}
\label{tab:fleiss_kappa}
\end{table}

\end{document}